\documentclass{emulateapj}

\def\gsim{\;\rlap{\lower 2.5pt
 \hbox{$\sim$}}\raise 1.5pt\hbox{$>$}\;}
\def\lsim{\;\rlap{\lower 2.5pt
   \hbox{$\sim$}}\raise 1.5pt\hbox{$<$}\;}
\def\entropyunits{$h_{70}^{-1/3}$ keV cm$^2$ }
\def\msolarunits{$h_{70}^{-1}$ $M_{\odot}$}

\begin{document}

\title{Cosmological Simulations of the Preheating Scenario for Galaxy Cluster Formation: Comparison to Analytic Models and Observations}
\author{Joshua D. Younger\altaffilmark{1,2} and Greg L. Bryan\altaffilmark{3}}
\altaffiltext{1}{Columbia Astrophysics Lab, Columbia Unviersity, Pupin Physics Laboratories, New York, NY 10027}
\altaffiltext{2}{Harvard--Smithsonian Center for Astrophysics, 60 Garden Street, 
Cambridge, MA 02138}
\altaffiltext{3}{Department of Astronomy \& Astrophysics, Columbia Unviersity, Pupin Physics Laboratories, New York, NY 10027}

\begin{abstract}

We perform a set of non--radiative cosmological simulations of a preheated intracluster medium in which the entropy of the gas was uniformly boosted at high redshift.  The results of these simulations are used first to test the current analytic techniques of preheating via entropy input in the smooth accretion limit.  When the unmodified profile is taken directly from simulations, we find that this model is in excellent agreement with the results of our simulations.  This suggests that preheated efficiently smoothes the accreted gas, and therefore a shift in the unmodified profile is a good approximation even with a realistic accretion history.  When we examine the simulation results in detail, we do not find strong evidence for entropy amplification, at least for the high-redshift preheating model adopted here.  In the second section of the paper, we compare the results of the preheating simulations to recent observations.  We show -- in agreement with previous work -- that for a reasonable amount of preheating, a satisfactory match can be found to the mass-temperature and luminosity-temperature relations. However -- as noted by previous authors -- we find that the entropy profiles of the simulated groups are much too flat compared to observations.  In particular, while rich clusters converge on the adiabatic self--similar scaling at large radius, no single value of the entropy input during preheating can simultaneously reproduce both the core and outer entropy levels.  As a result, we confirm that the simple preheating scenario for galaxy cluster formation, in which entropy is injected universally at high redshift, is inconsistent with observations.

\end{abstract}

\keywords{cosmology:theory -- intergalactic medium -- methods: numerical -- X--rays: galaxies: clusters}

\section{Introduction}

Clusters of galaxies are potentially powerful tools for precision cosmology.  In $\Lambda$CDM and other models of hierarchical structure formation, the mass function of clusters and its evolution are very sensitive to cosmological parameters \citep{bahcall1992,haiman2001,bahcallbode2003,younger2005}.  Recently, many authors have used the observed cluster abundance in a variety of wavelength ranges to constrain both the present day mean matter density $\Omega_m$ and the amplitude of linear fluctuations on 8 $h^{-1}$ Mpc scales \citep[e.g.,][]{henry1991,viana1996,borgani2001a,bahcall2003,pierpaoli2003,shuecker2003,henry2004,gladders2006}.  In the future, the first large-scale X--ray surveys will yield catalogs of many thousands of clusters \citep{haiman2005} that will place very tight constraints on cosmological parameters, including the dark energy equation of state and its evolution \citep{majumdar2004,wang2004,younger2006}.

With such a potentially promising dataset on the horizon, particular attention has been paid recently to addressing potential systematic uncertainties that may bias measurements of cosmology.  As \citet{henry2004} observed, the systematic uncertainty in determining $\sigma_8$ using current cluster measurements is driven by the calibration of the mass-temperature relation.  If cosmological results from future cluster surveys are to be robust, accurate theoretical models of cluster formation and evolution, including identification and proper treatment of the dominant heating and cooling mechanisms of the intracluster gas, are required.

The observational properties of the intracluster medium (ICM) are driven primarily by simple gravitational collapse.  If this were the only important physical mechanism, clusters would scale self--similarly \citep{kaiser1986}.  However, it has been known for some time that the observed bulk properties of galaxy clusters -- temperature, mass, luminosity -- do not conform to the self--similar model \citep{evrard1991,edge1991,david1993,markevich1998,arnaud1999,reiprich2002}.  This effect is also seen in the luminosity--temperature scaling of groups \citep{helsdon2000} and  the faintness of the unresolved $\sim 1$ keV X--ray background \citep{pen1999,wu2001,bryan2001}.  These departures from self--similarity are affected by non--gravitational baryonic physics; a complex interplay between various thermodynamic processes involving star formation and galaxy evolution \citep[see][and references therein]{rosati2002,voit2005}.

Motivated in part by observations suggesting a universal entropy floor for clusters \citep{ponman1999,lloyddavies2000}, many studies have proposed ``preheating'' of the gas in order to explain this departure from self--similarity \citep[e.g.,][]{kaiser1991,evrard1991,nfw95,cavaliere1997,balogh1999,ponman1999}.  In this scenario, the effects of energy input into the ICM from non--gravitational processes such as supernovae, star formation, and galactic winds are approximated by a high--redshift entropy modification that particularly affects low--mass systems \citep{bower1997,tozzi2001,voit2001,voit2002,voit2003}.  Subsequent numerical  \citep{borgani2001b,kay2004,borgani2005,muangwong2005} and analytic \citep{babul2002,voit2002,mccarthy2003a,mccarthy2004} modeling has been successful in reproducing many of the observed cluster scaling relations, both in the X--ray and Sunyaev--Zel'dovich (SZ) effect \citep{sunyaev1972,sunyaev1980}. 

In particular, \citet{voit2002,voit2003} propose that a simple shift of the entropy profile may be a good approximation to the effects of preheating.\footnote{Throughout this paper we use the term {\it preheating} to refer to a constant entropy increase, which corresponds to a spatially constant energy input at early times, when the gas is smooth.  \cite{voit2002} also discuss other forms of modifying the entropy profile, largely motivated by the effect of cooling and star formation, which we do not discuss in this paper.}  However, since this results was obtained in the smooth accretion limit, and real accretion of preheated gas is likely to be somewhat lumpy, it may not hold generally.  This is because it is not clear {\it a--priori} how efficiently universal preheating will smooth the gas distribution prior to cluster collapse.  In this work we compare the predictions of this shift model for preheating to the results of high--resolution cosmological simulations.  Such a comparison tests this analytic prescription, given a realistic accretion history.  And, it will motivate other model assumptions, such as the choice of boundary conditions, that are more difficult to derive from first principles.

At the same time, recent observations of the entropy profiles of nearby clusters have suggested that  preheating scenarios which invoke a universal entropy floor are incorrect in detail \citep{ponman2003,pratt2005,pratt2006}.  The same simulations we use to test analytic prescriptions offer the added benefit of a fully informed comparison of the predictions of the preheating scenario to observations.

This work is divided into three sections.  In \S~\ref{sec:sim}, we summarize the simulation and cluster identification procedure.  in \S~\ref{sec:model}, we compare the corresponding model predictions to the results of our simulations.  Finally, in \S~\ref{sec:obs} we compare the results of our simulations to recent observations of nearby clusters.  Throughout this work, ``preheating'' should be taken to refer specifically to the shift model of \citet{voit2002,voit2003}.

\section{The Simulations}

\label{sec:sim}
\subsection{Overview}
The simulations were performed with the Adaptive--Mesh--Refinement code {\sc enzo} \citep{bryan1999,norman1999,oshea2004} assuming a flat $\Lambda$CDM cosmology.  The cosmological model is consistent with the {\sc WMAP} Year 3 results \citep{spergel2006} with $(\Omega_m,\Omega_\Lambda,\Omega_b,h,\sigma_8) = (0.25,0.75,0.046,0.7,0.75)$.  We ran two complementary sets of simulations.  The first simulation was performed in a comoving periodic box with $L = 100$ $h^{-1}$ Mpc on a side with $N = 256^3$ dark matter particles and an equal number of grids, allowing for six levels of additional refinement that yielded a minimum cell size of $\sim 6$ $h^{-1}$ kpc.  This level of resolution was found (via a smaller set of higher resolution simulations) to be sufficient for reproducing the core entropy level of the smallest clusters we will investigate in this paper.  The initial conditions were generated with the fitting form of the linear dark matter power spectrum given by \citet{eisenstein1999} at an initial redshift of $z=60$ \citep[for details regarding the initialization of such simulations see][and references therein]{bertschinger1998}.

Due to its relatively small size, this 100 $h^{-1}$ Mpc simulation produced only a limited number of high mass clusters.  In order to fill out the higher--mass end of the scaling relations, we performed a second set of simulations using the same cosmology and method of generating initial conditions.  First,  a dark--matter only run was performed in a comoving periodic box $L = 400$ $h^{-1}$ Mpc on a side with $N = 128^3$ dark matter particles.  Then, the largest dark--matter halos were identified using the {\sc hop} algorithm \citep{eisenstein1998}, and the simulation was re--run with two nested grids: the first in the original $L = 400$ $h^{-1}$ Mpc box with $N = 128^3$ dark--matter particles and an equal number of grids, and the second in a high--resolution region, centered on one of the haloes, with $L \sim 50$ $h^{-1}$ Mpc and the same number of particles and grids, using the same refinement technique as the larger simulations.  This gave our high--resolution region equivalent mass and spatial resolution to the original runs.  We repeated this process for each of the four largest dark--matter haloes.

\subsection{Preheating the ICM}
\begin{figure*}
\plotone{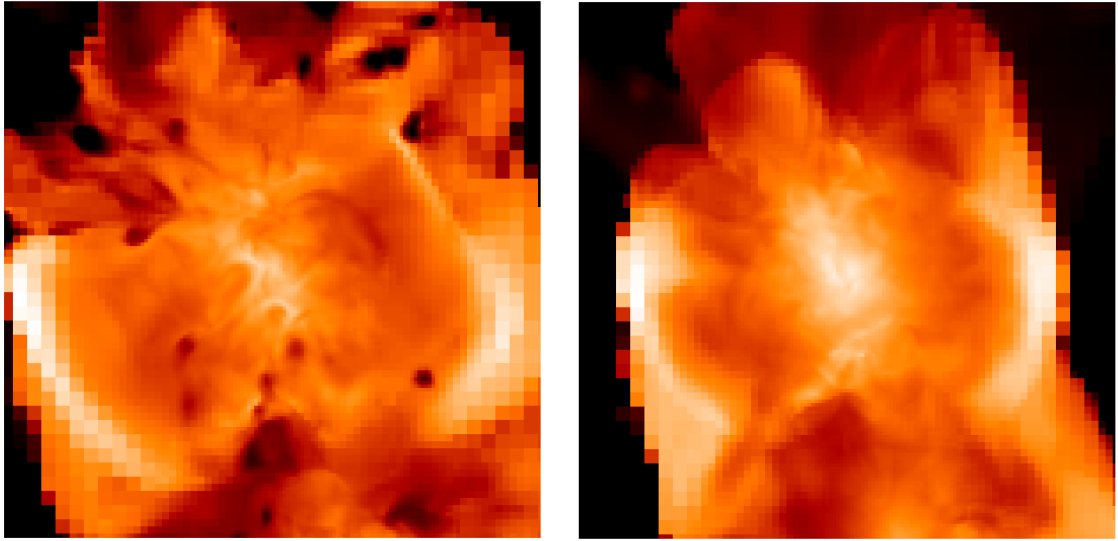}
\caption{Emission weighted 2--D projection of the $z=0$ gas temperature for a cluster with $M_{200} \sim 7\times 10^{14}$ \msolarunits\ two of the simulations: $K_0 = 0$ (left) and 311 \entropyunits (right).  Note how the preheating entropy input is effectively a smoothing operation.}
\label{fig:snapshot}
\end{figure*}

\citet{voit2003} have suggested that simple preheating in the smooth accretion limit is well--approximated by a universal shift of the unmodified cluster entropy profile.  This corresponds, in the context of a full cosmological simulation,  to raising the entropy level of all the gas in the universe at high--redshift, well in advance of cluster collapse, but within the time frame during which we expect to find galactic and stellar entropy input to the ICM.   Therefore, a fixed amount of entropy $K_0$ was added at $z = 10$ by increasing the thermal energy of each grid--point by
\begin{equation}
k\Delta T = K_0 \left (\frac{\rho_g}{m_p} \frac{2-Y_{\rm He}}{2}  \right)^{\gamma-1}
\end{equation}
where $\rho_g$ is the baryon density at that grid--point.  We assume an ideal gas consisting of a fully ionized H--He plasma with the cosmic helium mass fraction $Y_{\rm He}=0.25$ and $\gamma = 5/3$.  This was done for four cases: no preheating ($K_0 = 0$) and $K_0 = 78$, 155, and 311 \entropyunits.  The effects of radiative cooling and other additional sources of non--gravitational heating were neglected.  Snapshots of the $z=0$ emission--weighted 2--D projection of the gas temperature in the same cluster from two of the simulations are shown in Figure~\ref{fig:snapshot}.   This figure demonstrates visually that adding entropy smoothes the small-scale distribution, but leaves the larger scale structures largely intact.

\subsection{Cluster Identification and Analysis}
At each redshift output, clusters were identified using the {\sc hop} algorithm of \citet{eisenstein1998}, yielding clusters at $z = 0$ up to $M_{200} \sim 3\times 10^{14}$  \msolarunits\ in the 100 $h^{-1}$ Mpc simulations, and clusters up to $M_{200} \sim 8\times 10^{14}$ \msolarunits\ in the large box simulations, where $M_{200}$ is the total mass enclosed within a spherical overdensity of 200 relative to the critical density.  These were then analyzed using a \citet{raymondsmith1977} cooling function assuming a fully ionized H--He plasma with abundance $Z = 0.3 Z_{\odot}$.  The cluster center for large clusters ($M_{200} \gsim 4\times 10^{14}$ \msolarunits) was calculated using an iterative process of re--centering on the center of mass using successively smaller radii.  For all others, it was taken as the densest dark--matter particle from the {\sc hop} output.   We found that the choice of centering technique had no more than a 15\% effect on the core properties of the cluster, even at the highest masses.

For each cluster, we determined the mass ($T_{gw}$) and emission ($T_{ew}$) weighted gas temperatures, and the bolometric X--ray luminosity ($L_{bol}$).  In addition, we consider observables associated with the Sunyaev--Zel'dovich effect \citep[SZ;][]{sunyaev1972, sunyaev1980}, including the central decrement ($y_c$) and frequency independent, angular diameter distance scaled (FIADS) integrated SZ luminosity out to $R_{200}$ ($S_\nu d_A^2/f_\nu$).  These were calculated according to
\begin{equation}
y_c = \frac{\sigma_T}{m_ec^2}\int P_e d\ell,
\end{equation}
where $\sigma_T$ is the Thompson cross section, $m_e$ is the mass of the electron, $c$, is the speed of light, $P_e = n_e k T$ is the electron pressure, and $d\ell$ is the line element along the line of sight through the cluster center, and
\begin{equation}
\frac{S_\nu d_A^2}{f_\nu} =  \frac{(k T_{CMB})^3}{(h c)^2} \frac{\sigma_T}{m_ec^2}\int P_e dV,
\end{equation}
where $k$ and $h$ are  the Boltzmann and Planck constants respectively, $T_{CMB} = 2.728$ \citep{fixsen1996} is the Cosmic Microwave Background (CMB) temperature, and the integral is performed over a spherical volume out to $R_{200}$.  Finally, we calculated the gas--weighted entropy profile, where entropy was defined as
\begin{equation}
K_{gw} = k T_{gw} n_{e}^{1-\gamma}
\end{equation}
where  $n_e = \frac{\left < \rho_g\right >}{m_p} \frac{2-Y_{\rm He}}{2}$ is the average number density of electrons in a given shell, again assuming a fully ionized plasma.

\section{Comparison to Analytic Models}
\label{sec:model}
\subsection{The Analytic Model}

Our analytic model follows the techniques of \citet{voit2002} and \citet{younger2006}.  Given an initial, unmodified, spherically symmetric model in hydrostatic equilibrium, then entropy profile is ``modified'' -- in this case with a simple shift at all radii -- and the equations of hydrostatic equilibrium are re--integrated, yielding a final ``modified'' gas and temperature distribution.  Therefore, if $\hat{K}$ represents the shifted entropy profile, both the unmodified and modified models satisfy 
\begin{equation}
\frac{dP}{dr}  = \eta g(r) \rho_g(P,\hat{K}) 
\label{eq:hydro}
\end{equation}
where, casting the pressure and temperature in terms of $P$ and $\hat{K}$
\begin{equation}
\rho_g = \mu m_p \left[\frac{P}{k_B \hat{K}(M_g)}\right]^{3/5}.
\end{equation}
and assuming an ideal gas,
\begin{equation}
k_B T = \hat{K}(M_g)^{3/5}P^{2/5},
\end{equation}
keeping in mind $\hat{K}=K$ for the unmodified case. Furthermore, motivated in part by the simulations of \citet{dolag2005} and \citet{ascasibar2003}, we include $0 < \eta \leq1$ in (\ref{eq:hydro}) to allow for departures from strict hydrostatic equilibrium, in which the gas is supported by a combination of gas pressure and turbulent motions.  The gravitational potential of the dark matter (DM) follows a \citet{nfw97} (NFW) profile, as motivated by N-Body simulations, with a fixed concentration parameter $c=5$ \citep{eke2001}.  Finally, both the modified and unmodified models satisfy the same outer boundary condition, which is chosen to match the simulation results, but is related to the pressure required to resist the infall of baryonic matter at the viriol radius $P_{vir} \sim f_b \rho_{DM} v_{ff}^2$, where $f_b=\Omega_b/\Omega_m$ is the cosmic baryon baryon fraction.

The entropy floor is approximated by a shift in the entropy profile defined by:
\begin{equation}
\hat{K}(f_g) = K(f_g) + K_0.
\end{equation}
As discussed in \citet{voit2002}, this shift is a simple phenomenological approximation of a uniform, highredshift preheating.  In particular it assumes that mass shells remain in the same order after the modification (this is reasonable due to that fact that entropy must be monotonically increasing for stability reasons), and that the effect of preheating is to increase the entropy by the fixed, stated amount even after passing through the shock.  \citet{voit2002} examined this issue and found that this simple shift was a reasonable approximation (to within 20\%) for smooth accretion.  Of course, in a full cosmological simulation, the accretion is not generally smooth; however, as we will show, the shift approximation works remarkably well.

\subsection{Matching the Unmodified Case}

\begin{figure}
\plotone{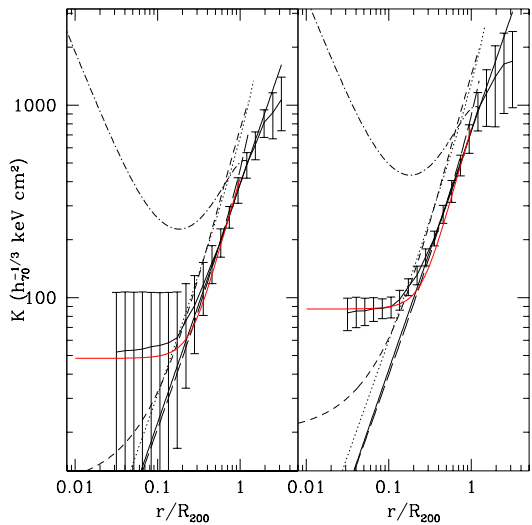}
\caption{Unmodified entropy profiles from different models as compared to the median profile from the simulations (solid, thick) for clusters in both a low (left; $4 < M_{200} < 6\times 10^{13}$ \msolarunits) and moderate (right; $1 < M_{200} < 1.5\times 10^{14}$ \msolarunits) mass bin along with error bars representing the $1-\sigma$ variance about the median, and the associated fit from \citet{voitkaybryan2005} (solid, thin).  The models shown; an NFW gas profile in hydrostatic equilibrium ($f_g = 0.8$, dotted), the isothermal model of \citet[][$f_g = 0.8$ $b = 0.8$, dashed]{makino1998}, gas with $\rho_g \sim (1+x)^{-3}$ ($f_g = 0.8$, dot--dash), and an NFW gas profile in modified hydrostatic equilibrium ($f_g = 0.9$ $\eta = 0.8$, long--dash).  The thick red line shows the fit to the simulation results used in our model.}
\label{fig:unmodified}
\end{figure}

For our comparison to be self--consistent, the most important consideration was our choice of unmodified distribution.  This must match the results of the no--preheating simulation as closely as possible; for if our unmodified model does not match the simulations without preheating, we have no reason to expect that modifications to that model will be consistent either.  We therefore attempted the match the median entropy profile in two mass bins: low ($4 < M_{200} < 6\times 10^{13}$ \msolarunits) and moderate ($1 < M_{200} < 1.5\times 10^{14}$ \msolarunits) mass clusters.    All of the model predictions were made at the median mass in each bin; $M_{200} = 5\times 10^{13}$ and $M_{200} = 1.3\times 10^{14}$ \msolarunits\ in the low and moderate mass bins respectively.  A summary of all the models considered is shown in Figure~\ref{fig:unmodified}.  

We first consider a set of {\it ab--initio} unmodified models.  Our first choice was similar to the fiducial model of \citet{voit2002} and \citet{younger2006}, in which the unmodified gas distribution is taken to follow the NFW dark matter density profile at the cosmic baryon density in strict hydrostatic equillibrium ($\eta = 1$).   Although it roughly reproduces the profile, in detail this NFW gas entropy profile was both too steep and incorrectly normalized, predicting a higher entropy for $r \gsim 0.1R_{200}$ than was seen in the simulations.  It furthermore did not reproduce the core entropy seen in simulations, due to the divergent NFW density at small radii.  Allowing $\eta$ to vary produced somewhat better agreement in the normalization and slope, but still predicted identically zero core entroy.

We then tried a set of alternative models in an attempt to reproduce the core entropy see in the simulations : isothermal and a gas distribution with a flat core.  Isothermal gas in hydrostatic equilibrium with $b = 0.8$ \citep[see][]{makino1998} also had incorrect slope and normalization, with a core entropy that was too low.  We then tried keeping the gas in hydrostatic equilibrium, but modifying the gas distribution such that $\rho_g \sim (1+x)^{-3}$.  This too, did not fit the simulations, requiring a divergent temperature at the cluster center and a non--monotonically increasing entropy profile, both of which were unphysical.

Since reproducing this median entropy profile from the no--preheating simulations with {\it ab--initio} models yielded no success, we instead chose to use a fit to the median entropy profile in the AMR simulations of \citet[]{voitkaybryan2005}\footnote{An alternative approach would have been to use polytropic fits to simulations, as in \citet{ostriker2005} and \citet{bode2006}.}.  We parameterized the entropy profile in terms of $f_g$, the enclosed gas fraction, with the following fit:
\begin{equation}
K(f_g)/K_{200} = 0.18 + 0.2 f_g + 1.5 f_g^2
\end{equation}
where $K_{200} = T_{200} (f_b 200 \rho_{\rm crit})^{-2/3} / (\mu m_p)$ and $T_{200} = GM_{200} \mu m_p/2 r_{200}$, following the definitions in \cite{voit2002}.   We chose the outer boundary condition, once again based on the simulations, to be given by $P_{200} = 0.7 T_{200} f_b \rho_{DM}(R_{200})$.

\citet{voitkaybryan2005} showed that the entropy profiles in SPH and AMR simulations agree very well outside of $0.2r_{200}$ while inside this radius, the SPH simulations show steeper entropy profiles with lower (but non-zero) core entropies.  The source of this discrepancy is not clear, although the new entropy conserving version of {\sc Gadget2} \citep{gadget2} does show higher entropy cores (G. Yepes, private communication).   In any case, the size of the difference at $r = 0.1r_{200}$ is quite small, about 25\% and certainly not enough to resolve the discrepancy with observations discussed in section~\ref{sec:obs}.

\begin{figure}
\plotone{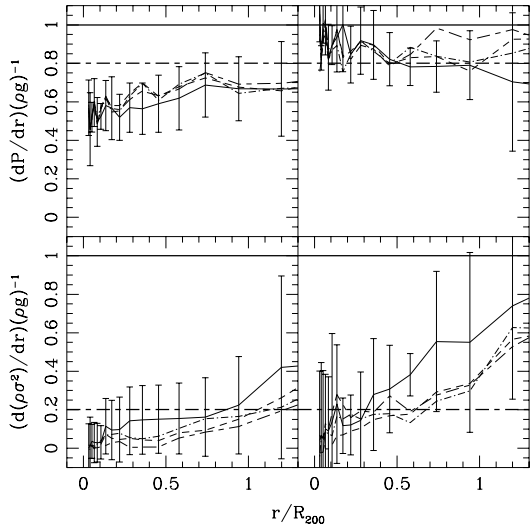}
\caption{Indicators of relative support from gas pressure (top) and turbulent motions (bottom) in the median profiles from the simulations for the same low (left) and moderate (right) mass bins from Figure~\ref{fig:unmodified}.  Shown are the results from four cases of injected entropy: no--preheating ($K_0 = 0$; solid) and $K_0 = 78$ (short--dash), 155 (dot--dash) and 311 (long--dash) \entropyunits.  The error bars representing the variance about the median profile are, for clarity, only shown for the no--preheating case, but are similar for the other three.  The thick solid line represent complete support against gravity, while the thick dashed line shows the relative contributions for $\eta = 0.8$.}
\label{fig:hydro}
\end{figure}

Finally, we note that our simulated clusters were not in strict hydrostatic equilibrium (see Figure~\ref{fig:hydro}), a result which is consistent with previous work \citep{ascasibar2003, dolag2005}.  The result of this was that even with our fitted entropy profiles, we could not reproduce the simulated density and temperature profiles in detail for $\eta=1.0$.  We found that setting $\eta$ to 0.8, as suggested by the median profiles in Figure~\ref{fig:hydro} (and previous work) generated a good fit to the gas profiles over a wide range of masses.

\subsection{Profiles}

\begin{figure}
\plotone{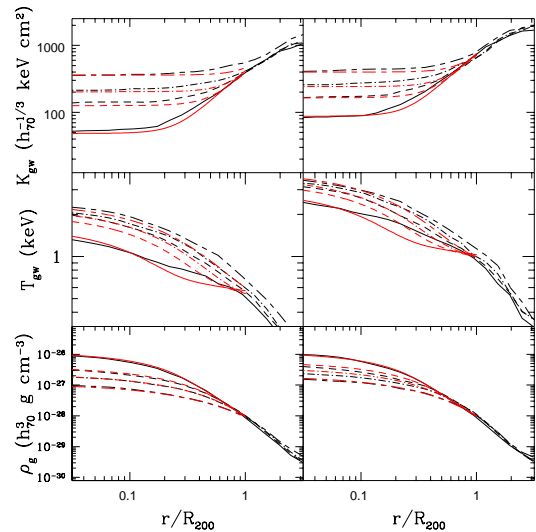}
\caption{Radial profiles at $z=0$ for the average gas density ($\rho_g$), gas weighted temperature ($T_{gw}$), and gas weighted entropy ($K_{gw}$) as a function of $r/R_{200}$ as predicted by both the numerical simulations (black, thick curve) and analytic model (red, thin curve) for four cases of injected entropy: no--preheating ($K_0 = 0$; solid) and $K_0 = 78$ (short--dash), 155 (dot--dash) and 311 (long--dash) \entropyunits.  The simulation results are the median profile for two representative mass ranges: $4 < M_{200} < 6\times 10^{13}$  (left) and $1 < M_{200} < 1.5\times 10^{14}$ (right) \msolarunits.  The analytic model predictions are taken at the median mass for each range in the numerical simulation: $5 \times 10^{13}$ (left) and $1.3\times 10^{14}$ (right) \msolarunits.}
\label{fig:profiles.model}
\end{figure}

Having found a reasonable entropy profile within our unmodified model, we then modied these profiles according to the prescription in \citet{voit2002} and \citet{younger2006} and computed the resulting model predictions for gas density, temperature, and entropy profiles in the same mass bins using matching entropy shifts of $K_0 = 0$ (no preheating), 78, 155, and 200 \entropyunits.  In Figure~\ref{fig:profiles.model}, we compare the model predictions to the simulation results for the gas density $\rho_g$, gas--weighted temperature $T_{gw}$, and gas--weighted entropy $K_{gw}$ profiles.  This comparison is made for the median profile in both the low and moderate mass bins.

We find remarkable agreement between our analytic predictions and the simulation results over both mass bins for all three entropy shifts.  The temperature profiles are underpredicted by the model, but never by more than $\sim  10\%$ at $R \lsim 0.1 R_{200}$, and never by more than $\sim 20\%$ out to $R_{200}$.   The entropy profile in the center is very well reproduced.  There is also a tendency for the model to underpredict the entropy at $R \gsim R_{500}$ compared to the simulations, although again never by more than about $20\%$.  

We can therefore conclude, as one can see qualitatively in Figure~\ref{fig:snapshot} that preheating efficiently smoothes the accreted density distribution, making the accretion histories of our simulated clusters more smooth.  As a result, shifting the unmodified entropy profile is a good approximation to the effects of preheating in the simulations.  Furthermore, we find that the model predictions and simulation results converge as $K_0$ increases.  This occurs because as $K_0$ increases, the entropy input can effectively smooth larger and larger sub--haloes, and the simulations converge on the smooth accretion limit, making our modified models a better and better approximation.

\subsection{The Intracluster Entropy}

From the comparison to the median profiles in Figure~\ref{fig:profiles.model}, we find both qualitative and quantitative agreement between the analytic predictions and the simulation results. This is very encouraging, however comparing the median profiles is only a snapshot of the behavior of the intracluster entropy for the median cluster profile in two mass bins.  The model predictions for the intracluster entropy over a much broader mass range contain a great deal of addition information.  

\begin{figure}
\plotone{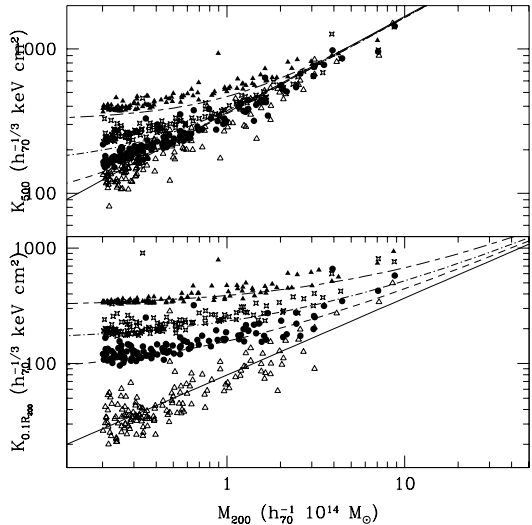}
\caption{The intracluster gas entropy scalings at $z=0$ as predicted as predicted by both the numerical simulations and analytic model (points/curves).  Shown are four different cases of injected entropy: no--preheating ($K_0 = 0$; open triangles/solid) and $K_0 = 78$ (filled circles/short--dash), 155 (stars/dot--dash) and 311 (filled triangles/long--dash) \entropyunits.  Show are the entropy at $0.1 R_{200}$ ($K_{0.1 R_{200}})$, and at $R_{500}$ ($K_{500}$) as a function of $M_{200}$.}
\label{fig:entropy.model}
\end{figure}

In Figure~\ref{fig:entropy.model}, we show the entropy--mass scalings for two characteristic entropy measures used in observations: the ``core'' entropy, taken at $r = 0.1 R_{200}$ (lower panel, $K_{0.1R_{200}}$), and the ``outer'' entropy taken at $r = R_{500}$ (upper panel, $K_{500}$).   For the core entropy, our analytic predictions are again in remarkable agreement with the simulations.  However at high mass $M_{200} \gsim 3\times 10^{14}$ \msolarunits, our analytic predictions are somewhat lower than the simulation results.  This behavior is slightly enhanced for higher values of $K_0$.  Conversely, the models somewhat underpredict the outer entropy at all masses, with the exception of the very highest mass clusters where small numbers of objects make the trend difficult to discern.

The agreement we see for the entropy of the preheated models is of particular interest and somewhat surprising.
The effects of preheating on the entropy of simulated clusters in the cosmological simulation represents contributions from two competing effects.  Entropy input into the ICM will tend to push out the accretion shock, which will in turn decrease the efficiency of heating at the shock and decrease the core entropy \citep{balogh1999,tozzi2001,voit2003}.  However, as first suggested by \citet{voit2003}, preheating will also tend to lower the density of accreted haloes.   This may affect a partial transition from lumpy to smooth accretion that will actually increase the efficiency of entropy generation at the accretion shock, and with it the entropy of the system.  Our simulations suggest that these competing effects tend to cancel each other out.

\begin{figure}
\plotone{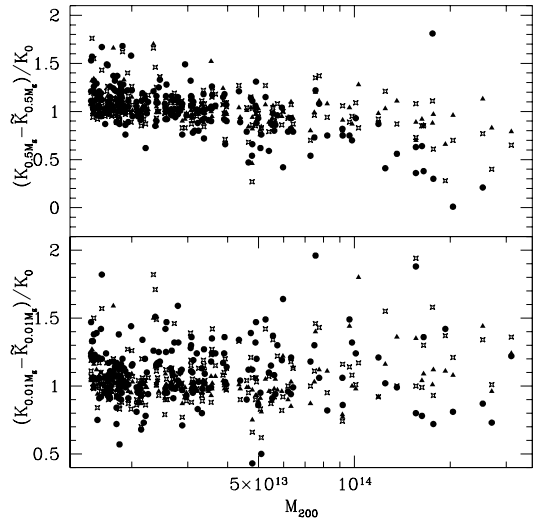}
\caption{Same as Figure~\ref{fig:entropy.model}, for $(K_{0.01M_g}-\tilde{K}_{0.01M_g})/K_0$ and $(K_{0.5M_g}-\tilde{K}_{0.5M_g})/K_0$, where $K_{0.01M_g}$ is the entropy at fixed gas fraction $0.01M_g$, $K_{0.5M_g}$ is the entropy at fixed gas fraction $0.5M_g$, $\tilde{K}$ represents the no--preheating case, and $M_g$ is the total gas mass at $R_{200}$.  Note that a value of $1$ represents no entropy amplification.}
\label{fig:entropyamp}
\end{figure}

In order to extract only the effects of entropy input on the simulations, we match the halos in each simulation, and do a halo--by--halo comparison. For each cluster, we examine two ratios, or amplification factors: $f_{core} = (K_{0.01M_g}-\tilde{K}_{0.01M_g})/K_0$ and $f_{outer} = (K_{0.5M_g}-\tilde{K}_{0.5M_g})/K_0$, where $K_{0.01M_g}$ is the gas--weighted average entropy $\left<T n_e^{-2/3}\right>$ at fixed gas fraction $0.01M_g$, $K_{0.5M_g}$ is the entropy at fixed gas fraction $0.5M_g$, and $M_g$ is the total enclosed gas mass at $R_{200}$.  We use $\tilde{K}$ to indicate the no--preheating case, and compute the ratio for each of the preheated simulations.  These ratios represent the entropy offset between simulated clusters at fixed gas mass, in units of the entropy input $K_0$, at fixed gas mass in both the core and outskirts.  The results of this comparison are show in Figure~\ref{fig:entropyamp}, with $f_{core}$ or $f_{outer} = 1$ representing no entropy amplification, and larger values of $f$ implying more entropy amplification.

We find only mild evidence of entropy amplification in Figure~\ref{fig:entropyamp}.  The overall mean core and outer amplification factors for $K_0 = (78,155,311)$ \entropyunits are $\left<f_{core}\right> = (1.13,1.09,1.03)$ and $\left<f_{outer}\right> = (0.99,1.03,1.02)$ with associated variances $\sigma_{core} = (0.36,0.36,0.18)$ and $\sigma_{outer} = (0.34,0.21,0.16)$.  The trend towards more amplification at lower masses for both amplification factors is qualitatively consistent with \citet{voit2003} and \citet{borgani2005}.  However, the mean amplitude of the amplification is $\lsim 15\%$ for all values of $K_0$, and within a wide scatter is consistent with unity.  Overall we find that on average, preheating does make entropy generation more efficient, but only by a small fraction.

\citet{borgani2005} found amplification factors of $\sim 1.14$ and 1.84 at $R = 0.1R_{vir}$ for a simulated group of mass $1.64 \times 10^{13}$ \msolarunits\ using entropy floors of $\sim 25$ and 100 \entropyunits respectively.  As noted above, we find smaller values for a much larger sample.  One possible source of this discrepancy is the nature of the simulations: \citet{borgani2005} include the effects of heating/cooling from a uniform UV background, star formation from a multiphase interstellar medium, and galactic winds powered by supernova feedback, whereas our simulations use a simple implementation of preheating in which all of these processes are subsumed into the high redshift entropy modification.  Another possibility is the improved statistical power of our simulations, which model more objects than \citet{borgani2005}.  Although there are groups of comparable mass in our sample with amplification factors that are roughly consistent with the findings of \citet{borgani2005}, we also find that the mean amplification is within $1-\sigma$ of unity, even for low--mass systems.  However, perhaps the most likely explanation is the timing of the entropy input.\footnote{We thank Mark Voit for pointing out this possibility.}  In \citet{borgani2005}, the entropy was added much later, at $z=3$, when many smaller halos would have already formed.  The gas in these halos is at high density and so, for a fixed amount of entropy, receives much more energy.  This may have lead to large-scale outflows and shocking, which, in turn, may have caused the entropy amplification.  In the simulations discussed here, the input is done at much earlier times ($z=10$) as in \citet{bialek2001}, because we want to isolate only the effects of preheating.

We also find some evidence of turbulent mixing in the gas.  This will will tend to equalize the entropy profile, raising the core and lowering the outer entropy.  It has previously been noted that simulated clusters are not in strict hydrostatic equilibrium \citep{ascasibar2003}, probably due to turbulence \citep{dolag2005}.  Figure~\ref{fig:entropyamp} exhibits two telling trends: the core amplification factor $f_{core}$ tends towards unity with higher entropy input $K_0$, and the outer amplification factor $f_{outer}$ tends toward entropy suppression ($f_{core} < 1$) in systems with $M_{200} > 5\times 10^{13}$ \msolarunits.  This is consistent with turbulent mixing, which will be less efficient at amplifying the core entropy at higher $K_0$ as smoothing suppresses the turbulence, and more efficient at suppressing the outer entropy at higher mass as more of the substructure is resolved and turbulence becomes more important.  These two tends are, as we show in Figure~\ref{fig:hydro}, borne out in our simulations; higher mass clusters with higher $K_0$ tend to have more support against gravity contributed by turbulent motions in the gas.  However, though we see trends that suggest turbulent mixing, we concede that this is speculation and warrants a more in depth analysis, which we defer to future work.

Therefore, our simulations show the combined effect of two processes that are difficult to disentangle.  Classic entropy amplification will tend to increase both the core and outer entropy, and will tend towards higher $f_{core}$ at lower mass.  Turbulent mixing will suppress the outer entropy, tending towards lower $f_{outer}$ at higher mass, while simultaneously raising the core entropy more efficiently at higher $K_0$.  We see all of these trends in our data.  As a result, while the magnitude of the amplification favors the more subtle effects of turbulent mixing, we cannot firmly argue that either is dominant.  Nevertheless, as Figure~\ref{fig:entropyamp} makes clear, the size of either effect is not large.

\begin{figure}
\plotone{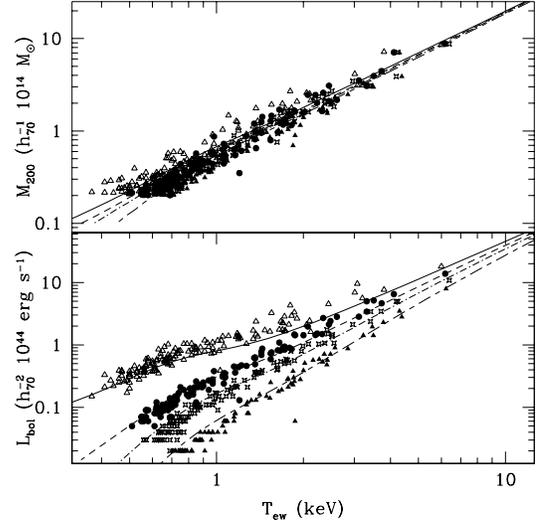}
\caption{Same as Figure~\ref{fig:entropy.model}, but for the temperature--luminosity and mass--temperature scalings.The luminosity $L_{bol}$ is the bolometric X--ray luminosity and $T$ is the emission weighted gas temperature.}
\label{fig:scaling.model}
\end{figure}

\subsection{Scaling Relations}

\begin{figure}
\plotone{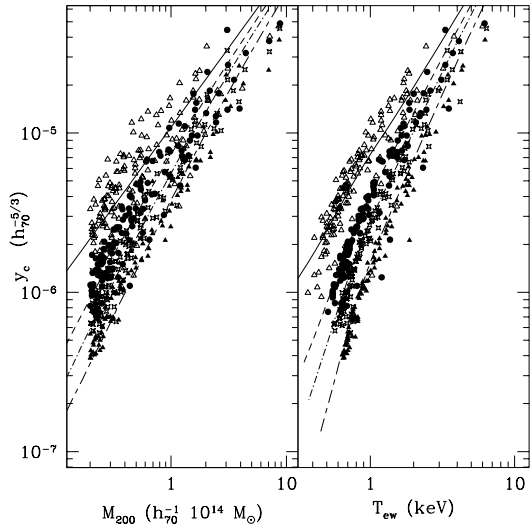}
\caption{Same as Figure~\ref{fig:scaling.model} , for the central SZ decrement $y_c$.}
\label{fig:yc.model}
\end{figure}

\begin{figure}
\plotone{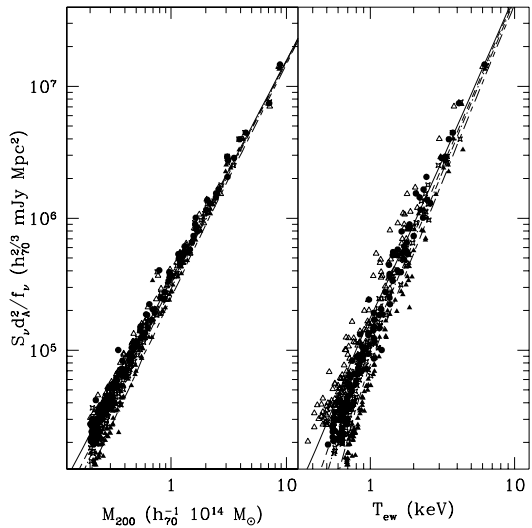}
\caption{Same as Figure~\ref{fig:scaling.model}, for the frequency independent, angular diameter distance scaled integrated SZ luminosity $S_{\nu} d_A^2/f_{\nu}$.}
\label{fig:yint.model}
\end{figure}

In addition to the entropy scalings, we compare the model predictions to the simulation results for several observationally motivitated scaling relations.  In Figure~\ref{fig:scaling.model}, we present two scalings typically used by X--ray observers: the mass--temperature (for $M_{200}$), and temperature--luminosity (for $L_{bol}$, the bolometric X--ray luminosity) relations.  In Figures~\ref{fig:yc.model} and \ref{fig:yint.model} respectively, we present the mass and temperature scalings for two SZ observables defined above: the central decrement $y_c$ and the FIADS SZ luminosity $S_{\nu} d_A^2/f_\nu$.  

For the typical X--ray observables show in Figure~\ref{fig:scaling.model}, we again find remarkable agreement between our model and the simulations.  This is due largely to the agreement between the predicted and simulated core density, as the X--ray emission scales as $\sim \rho^2$ and is only weakly dependent on the temperature of the gas.  Furthermore, this agreement suggests that the scalings in Figure~\ref{fig:scaling.model} are largely insensitive to asymmetries in the simulations; we find that our assumption of spherical symmetry is a good approximation to the simulation results.

In Figures~\ref{fig:yc.model} and \ref{fig:yint.model}, we present the mass (left panel) and temperature (right panel) scaling of two typical SZ observables.   Our analytic predictions are in good agreement with the simulation results for both scalings.  We furthermore find that the central SZ decrement, in both the model and simulations, to be a far better indicator of cluster temperature than mass.  We also confirm previous studies, which found the FIADS SZ integrated luminosity to be a robust, low--scatter mass and temperature indicator with roughly self--similar scaling \citep{dasilva2000,motl2005,nagai2006}.   This mass--FIADS SZ luminosity scaling is remarkably independent of the value of the entropy input $K_0$, while the temperature scaling exhibits weak $K_0$ dependence.

\subsection{The Importance of the Core Entropy}

We have found very good agreement between our analytic model predictions and the simulation results for the median density, temperature, and entropy profiles, the core and outer entropy scalings, and six different observable scalings.  It is important to note, however, that this agreement came in large part by imposing an unmodified entropy profile derived from the simulations.  None of our {\it ab--initio} unmodified models were in agreement with the simulation results.

In particular, even when the slope and normalization of the entropy profile were matched closely, as in the case of an NFW gas profile with $\eta = 0.8$ (see Figure~\ref{fig:unmodified}), there was poor agreement with the simulation results.  This was true even of the scaling of integrated quantities such as the $L_{bol}-T$ and $M_{200}-T$ relations, which were too steep and incorrectly normalized.  At the same time, as the value of the entropy input $K_0$ was increased, we found improved agreement with the simulations.  

This convergence is consistent with what we showed explicitly earlier; preheating efficiently smoothes the ICM, and therefore a simple shift of the unmodified entropy profile is a good approximation for the effects of preheating \citep{voit2003}.  At the same time, an unmodified model with no core entropy was quantitatively and qualitatively at odds with simulation results, even when that unmodified entropy profile was a very good fit for $r \gsim 0.1 R_{200}$.  And, though the value of this core entropy can vary depending on the numerical technique used \citep{voitkaybryan2005,oshea2005}, it is essential in setting the thermodynamic state and structural properties of the ICM.  The importance of this non--zero core entropy has been previously downplayed \citep{bryan2000,voit2002,voitkaybryan2005}, but here we find it to be a primary cause of disagreement between analytic model predictions and hydrodynamical simulations, even when the modification procedure is shown to be a good approximation.

\section{Comparison to Observations}
\label{sec:obs}

\begin{figure}
\plotone{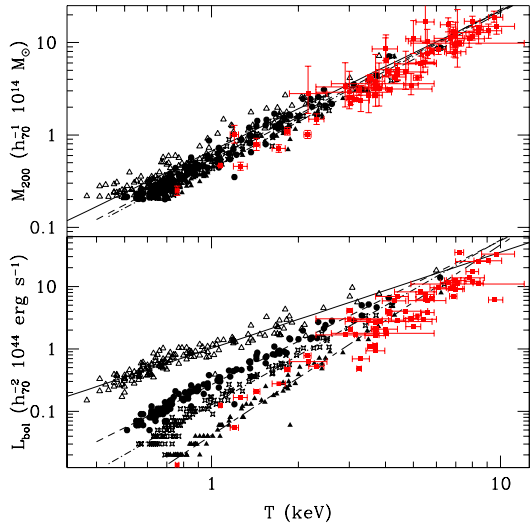}
\caption{Temperature--luminosity and mass--temperature scalings at $z=0$ as predicted by both the numerical simulations with their associated power--law fits (black, points/curves).  Shown are four different cases of injected entropy: no--preheating ($K_0 = 0$; open triangles/solid) and $K_0 = 78$ (filled circles/short--dash), 155 (stars/dot--dash) and 311 (filled triangles/long--dash) \entropyunits.  Also plotted are the observations of \citet[][red filled squares]{reiprich2002}.}
\label{fig:scaling.obs}
\end{figure}

\begin{figure}
\plotone{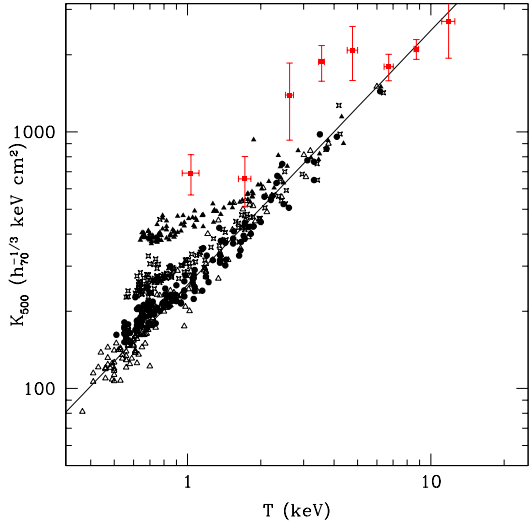}
\caption{Same as Figure~\ref{fig:scaling.obs}, but for the outer cluster entropy $K_{500}$.  Also show are the binned observations of \citet[][red filled squares]{ponman2003}.  The solid line is the self--similar prediction from the simulations.  The solid line is a power--law fit to the self--similar prediction from the simulations.}
\label{fig:outer.entropy.obs}
\end{figure}

\begin{figure}
\plotone{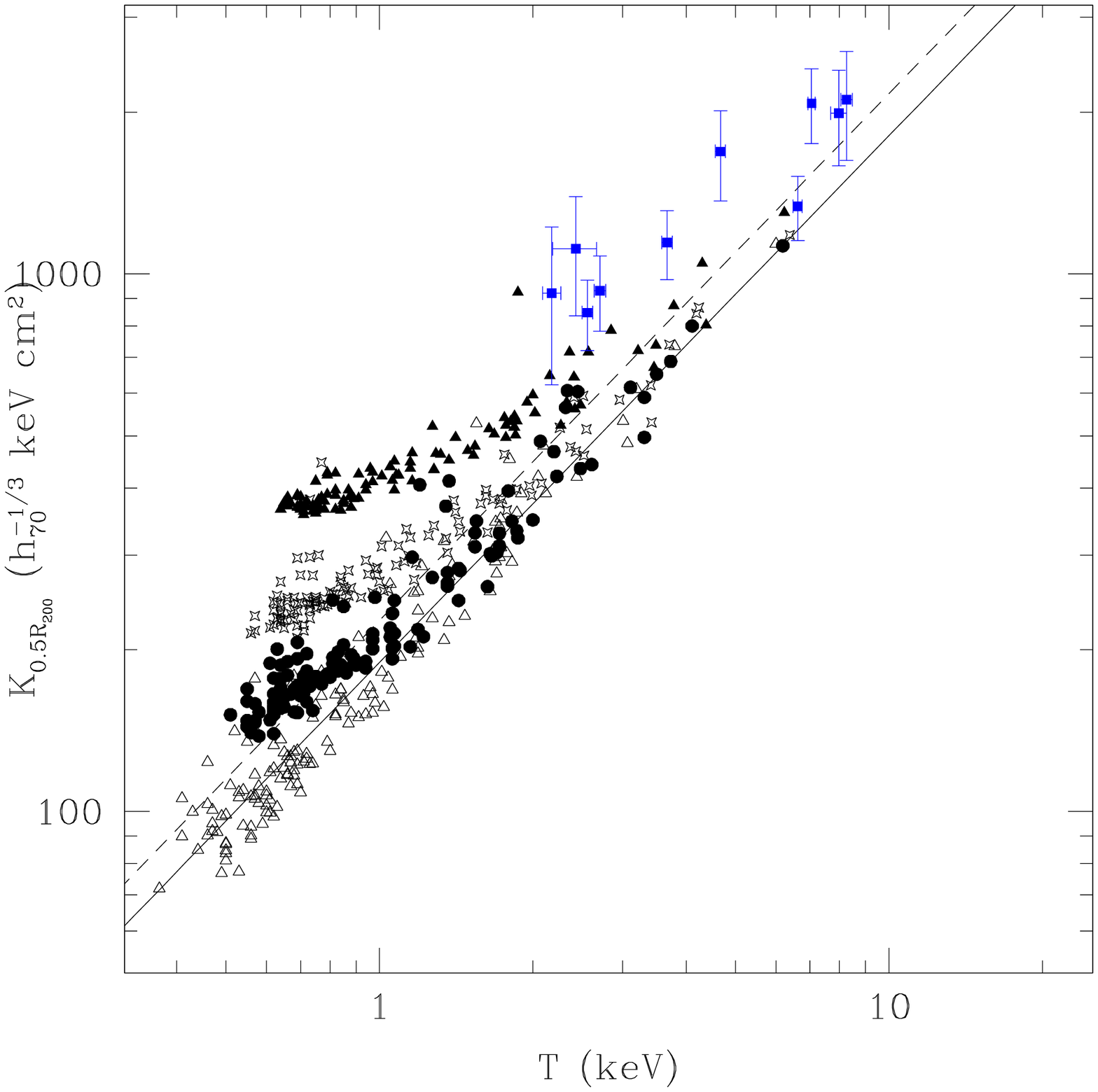}
\caption{Same as Figure~\ref{fig:scaling.obs}, but for the outer cluster entropy $K_{0.5R_{200}}$ to match the observations, which is similar to $K_{500}$.  Also show are the observations of \citet[][blue filled squares]{pratt2006}.  The solid line is a power--law fit to the self--similar prediction from the simulations.  The dashed line is the same relation, approximately corrected to the baryon fraction $f_b$ used in \citet{pratt2006} for comparison to the adiabatic scaling.  We also note that, for three of the observed clusters, the measurement presented here is an extrapolation from $\sim0.3R_{200}$ and may be uncertain.}
\label{fig:outer.entropy.obs2}
\end{figure}

\begin{figure}
\plotone{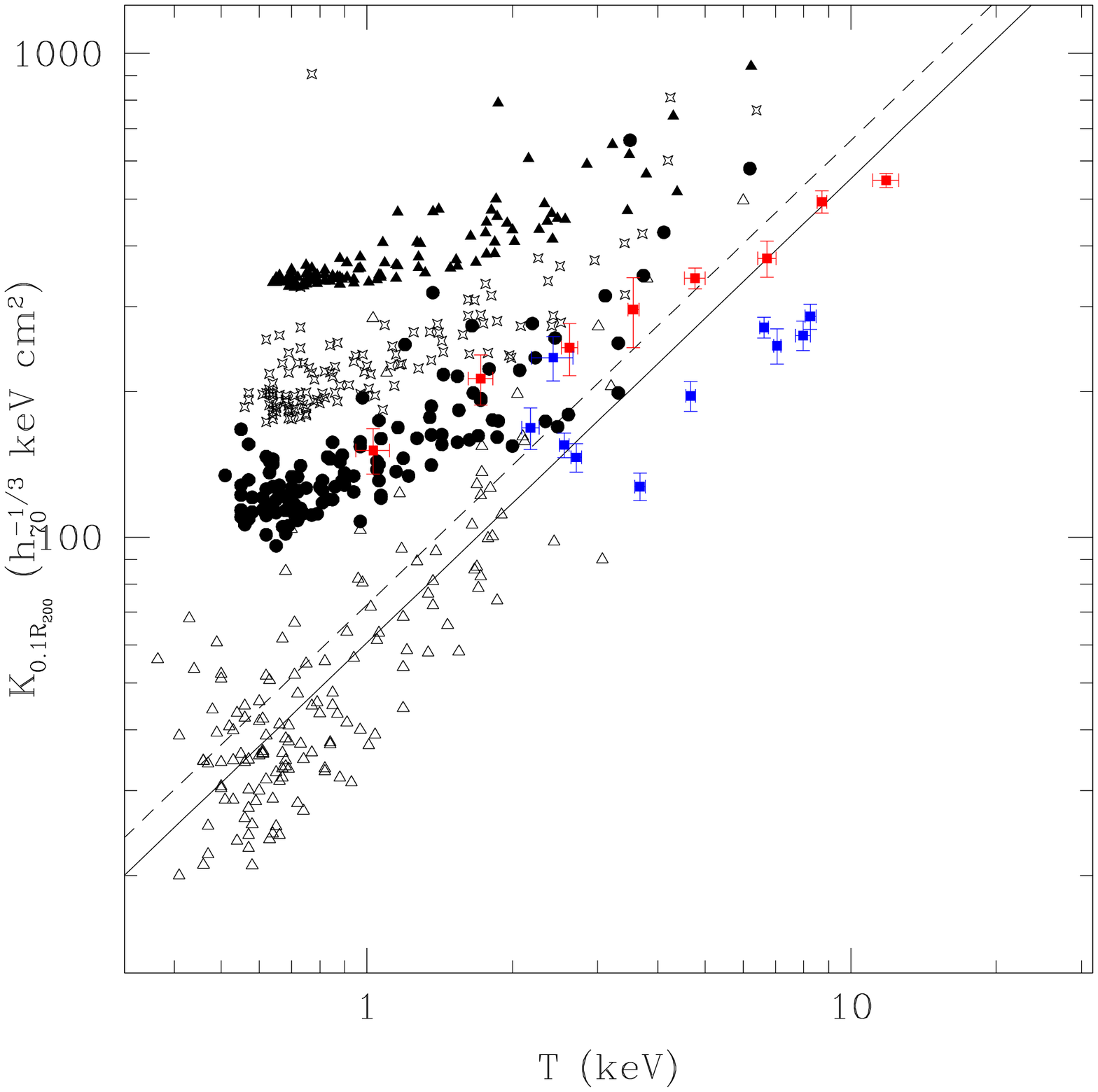}
\caption{Same as Figure~\ref{fig:outer.entropy.obs}, but for the core cluster entropy $K_{0.1 R_{200}}$.  Also show are the binned observations of \citet[][red filled squares]{ponman2003} and \citet[][blue filled squares]{pratt2006}.  The solid line is a power--law fit to the self--similar prediction from the simulations.    The dashed line is the same relation, approximately corrected to the baryon fraction $f_b$ used in \citet{pratt2006} for comparison to the adiabatic scaling.  The offset between the two observations is likely due to the fact that \citet{pratt2006} resolve the temperature profiles of their clusters, whereas \citet{ponman2003} do not.  This may lead to an over--estimation of the core entropy by \citet{ponman2003}}
\label{fig:core.entropy.obs}
\end{figure}

There has been some suggestion in the literature, motivated by recent observations, that preheating models which invoke a universal entropy floor are incorrect in detail \citep{ponman2003,pratt2005,pratt2006}.  In particular, some authors have suggested that this brand of preheating predicts isentropic cores in low--mass systems that are not observed.  In this section, we use the results of our simulations, which necessarily include all the relevant dynamical and geometric effects\footnote{We note as a caveat that our simluations neglect radiative cooling, which may erode isentropic cores in low--mass clusters.},  as a test of the preheating scenario itself.   Towards this end, we compare our results to observations of X--ray scalings (Figure~\ref{fig:scaling.obs}) and both the core and outer entropy--temperature  scalings (Figures~\ref{fig:outer.entropy.obs}, \ref{fig:outer.entropy.obs2}, and \ref{fig:core.entropy.obs}). j

In Figure~\ref{fig:scaling.obs}, we show the simulation results, along with best--fit power--laws, for the mass--temperature (upper panel) and temperature--luminosity (lower panel) scaling relations as compared to the observations of \citet{reiprich2002}.  As we mention above, the mass--temperature relation from the simulation is roughly independent of the value of the entropy input $K_0$.  And, not surprisingly, with the exception of the no--preheating simulation all of the simulated mass--temperature relations are consistent with the observational data.  The temperature--luminosity relation, on  the other hand, is very sensitive to the entropy input, as the bolometric luminosity scales roughly as $L_{bol} \sim \rho_g^2$ and preheating depresses and flattens the core density distribution.    The temperature and luminosity are also both pure observables given a cosmological model -- as opposed to $M_{200}$ which \citet{reiprich2002} estimate using an isothermal $\beta$--model \citep{cavaliere1976} -- and therefore provide more robust constraints on $K_0$.  We find that an entropy input $155 \lsim K_0 \lsim 311$ \entropyunits will best fit the observational data.  

Figures~\ref{fig:outer.entropy.obs}, \ref{fig:outer.entropy.obs2}, and \ref{fig:core.entropy.obs} show the outer ($K_{500}$ and $K_{0.5R_{200}}$) and core ($K_{0.1 R_{200}}$) entropy--temperature scaling from the simulations as compared to the observations of \citet{ponman2003} and \citet{pratt2006}.   Our simulation results show the nearly isentropic cores in low--mass (or temperature) clusters: notice that the entropy is nearly mass--independent for low-mass clusters at the highest level of entropy input.  In addition there is a break in the power-law scaling which shifts to higher mass with higher entropy input.  

We find that, while observations of the entropy at large radius in rich clusters are consistent our results, fully three--dimensional, non--radiative cosmological simulations of the preheating scenario of cluster formation are inconsistent with observations in several ways.  First, the entropy input scaling suggested by the temperature--luminosity relation is clearly inconsistent with observations of the intracluster entropy; simulations of the best fit $K_0$ range overproduce the core entropy and underproduce the outer entropy at fixed temperature.  Second, the entropy--temperature scalings are themselves inconsistent as compared to the observations; the best--fit entropy input value as determined by the core entropy underproduces the outer entropy, and vice--versa.  Finally, we show the insentropic low--mass cores predicted by preheating as a function of $K_0$, which are not observed.

As a caveat to our comparison to observations, we note that there has been some suggestion, most notably by \citet{mccarthy2004} that observational studies of the entropy profiles of clusters have been biased towards low core entropy systems.  This may, they argue, explain why some observations appear to be inconsistent with high levels of entropy input \citep{ponman2003,mccarthy2004,pratt2005,pratt2006}.  However, while this is a possibility, we find dramatic and qualitative disagreement between the scalings in our simulations simulations and observations which would be difficult to reproduce with a low--core entropy selection bias.  We furthermore find disagreement at large radius (see Figures~\ref{fig:outer.entropy.obs} and \ref{fig:outer.entropy.obs2}), well outside the cool region.  It is unlikely that a bias related to low entropy in the core would affect the outskirts.

\subsection{Including Radiative Cooling}

As an aside, we briefly investigate the effects of radiative cooling.  There has been some hope in the literature that the shortcomings of the preheating scenario will be solved by a combination of preheating and radiative cooling \citep{voit2002,mccarthy2004}.  Unfortunately this is difficult to implement in a simulation because heating and cooling are probably strongly tied together \citep[e.g.,][]{kay2004, muangwong2005}.  Nevertheless, we can imagine one simple and concrete extension to the straightforward preheating picture in which we include radiative cooling (but not star formation or feedback) at some late epoch, well after preheating has occurred.  We therefore re--ran our $K_0 = 155$ \entropyunits simulation, including the effects of radiative cooling for $z < 3$ \citep[see][]{oshea2004}.  We find that although the core entropy is modified, the entropy at $R_{500}$ is relatively insensitive to cooling and so the basic results above do not change.  This is, however, another topic worthy of much more thorough investigation.

\section{Conclusion}
We present high--resolution hydrodynamic simulations of the preheating scenario of cluster formation, in which the effect of SN and/or AGN heating is approximated as a universal increase of the entropy, implemented at high redshift.  We focus on two aspects: (1) how well analytic models of entropy input reproduce the simulations results, and (2) how well the preheating simulations match current cluster observations.

We find that analytic models of preheating following those of \citet{voit2002} and \citet{younger2006} are remarkably successful at reproducing the results of hydrodynamical simulations in detail.  This agreement extends from scalings of integrated quantities such $L_{bol}-T$, $M_{200}-T$, and SZ scalings for $y_c$ and the FIADS integrated SZ luminosity, to core and outer entropy scalings, and finally to the median entropy, temperature, and density profiles of both low and moderate mass clusters.  However, this agreement is dependent on the right choice of an unmodified profile; in particular one with non--zero core entropy.

When we look in detail at the simulation results, we do not find strong evidence for entropy amplification, the process in which the entropy is amplified even beyond the level injected \citep{voit2003}.  This differs from the results of \citet{borgani2005}, who looked at a small set of clusters and groups.  The reasons for this are not entirely clear although their preheating model differed from that used here (in particular the entropy input occured at a lower redshift).

Our simulations also show that the preheating scenario for cluster formation is in conflict with recent observations of the X--ray scalings \citep{reiprich2002} and intracluster entropy \citep{
ponman2003,pratt2005,pratt2006}.  The best--fit value of the entropy input $K_0$, determined from the X--ray scalings (see Figure~\ref{fig:scaling.obs}), underproduces the entropy at $R_{500}$ and overproduces the entropy at $0.1R_{200}$ at fixed temperature (see Figures~\ref{fig:outer.entropy.obs} and \ref{fig:core.entropy.obs}).   No single value for the entropy input due to preheating can simultaneously match the observed entropy values at both the core {\it and} the outskirts.  This result is in qualitative agreement with previous work \citep{borgani2005}.  The inclusion of radiative cooling at late times (without any additional feedback) does not appear to change this result.  Therefore, we find this simple preheating scenario to be in disagreement with observations, implying that other, more sophisticated treatments are needed.

\acknowledgements
We thank Zolt\'{a}n Haiman, Stefano Borgani, Mark Voit and Sheng Wang for helpful comments. In addition, we thank the referee for helpful suggestions.  This work was supported in part by NSF grant AST-05-07161, by the U.S. Department of Energy under Contract No. DE--AC02--98CH10886, and by the Columbia University Initiatives in Science and Engineering (ISE) funds.  Greg Bryan acknowledges support from NSF grants AST-0547823, and AST-0606959.


\begin{thebibliography}{}

\bibitem[Arnaud \& Evrard(1999)]{arnaud1999} Arnaud, M. \& Evrard, A. E., 1999, MNRAS, 305, 631
\bibitem[Ascasibar et al.(2003)]{ascasibar2003} Ascasibar, Y., Yepes, G., Mueller, V., \& Gottloeber,, S., 2003, MNRAS, 346, 731
\bibitem[Babul et al.(2002)]{babul2002} Babul, A., Balogh, M. L., Lewis, G. F., \& Poole, G. B., 2002, MNRAS, 330, 329
\bibitem [Balogh, Babul, \& Patton(1999)]{balogh1999} Balogh, M. L., Babul, A., \& Patton, D. R., 1999, MNRAS, 307, 463
\bibitem[Bahcall \& Bode(2003)]{bahcallbode2003} Bahcall, N. A. \& Bode, P., 2003, ApJ, 588, L1
\bibitem[Bahcall \& Cen(1992)]{bahcall1992} Bahcall, N. A. \& Cen, R., 1992, ApJ, 398, L81
\bibitem[Bahcall et al.(2003)]{bahcall2003} Bahcall, N., et al., 2003, ApJ, 585, 182
\bibitem[Balogh et al.(2006)]{balogh2006} Balogh, M. L., Babul, A., Voit, M. G., McCarthy, I. G., Jones, L. R., Lewis, G. F., \& Ebeling, H., 2006, MNRAS, 366, 624
\bibitem[Bertschinger(1998)]{bertschinger1998} Bertschinger, E., 1998, ARA\&A, 36, 599
\bibitem[Bialek, Evrard, \& Mohr(2001)]{bialek2001} Bialek, J. J., Evrard, A. E., \& Mohr, J. J., 2001, ApJ, 555, 597
\bibitem[Bode et al.(2006)]{bode2006} Bode, P., Ostriker, J.~P., Weller, J., \& Shaw, L., 2006, ApJ, in press, astro-ph/0612663 
\bibitem[Borgani et al.(2001a)]{borgani2001a} Borgani, S., Rosati, R., Tozzi, P., Stanford, S. A., Eisenhardt, P. R., Lidman, C., Holden, B., Della Ceca, R., Norman, C., \& Squires, G., 2001a, ApJ, 561, 13
\bibitem[Borgani et al.(2001b)]{borgani2001b} Borgani, S., Governato, F., Wadsley, J., Menci, N., Tozzi, P., Lake, G., Quinn, T., \& Stadel, C., 2001b, ApJL, 559, 71
\bibitem[Borgani et al.(2005)]{borgani2005} Borgani, S., Finoguenov, A., Kay, S. T., Ponman, T. J., Springel, V., Tozzi, P., \& Voit, M. G., 2005, MNRAS, 361, 233
\bibitem[Bower(1997)]{bower1997} Bower, R. G., 1997, MNRAS, 288, 355
\bibitem[Bryan(2000)]{bryan2000} Bryan, G. L. 2000, ApJ
\bibitem[Bryan \& Voit(2001)]{bryan2001} Bryan, G. L. \& Voit, G. M., 2001, ApJ, 556, 590
\bibitem[Bryan(1999)]{bryan1999} Bryan, G. L., 1999, Comput. Sci. Eng., 1, 46
\bibitem[Cavaliere \& Fusco--Femiano(1976)]{cavaliere1976} Cavaliere, A. \& Fusco--Femiano, R., 1976, A\&A, 49, 137
\bibitem[Cavaliere, Menci, \& Tozzi(1997)]{cavaliere1997} Cavaliere, A., Menci, N., \& Tozzi, P., 1997, ApJ, 484, L21
\bibitem[David et al.(1993)]{david1993} David, L. P., et al., 1993, ApJ, 412, 479
\bibitem[da Silva et al.(2000)]{dasilva2000} da Silva, A. C., Kay, S. T., Liddle, A. R., \& Thomas, P. A. 2004, MNRAS, 348, 1401
\bibitem[Dolag et al.(2005)]{dolag2005} Dolag, K., Vazza, F., Brunetti, G., \& Tormen, G., 2005, MNRAS, 364, 753
\bibitem[Edge \& Stewart(1991)]{edge1991} Edge, A. C. \& Stewart, G. C., 1991, MNRAS, 252, 414
\bibitem[Eisenstein \& Hu(1999)]{eisenstein1999} Eisenstein, D. J. \& Hu, W., 1999, ApJ, 511, 5
\bibitem[Eisenstein \& Hut(1998)]{eisenstein1998} Eisenstein, D. J. \& Hut, P., 1998, ApJ, 498, 137
\bibitem[Eke, Navarro, \& Steinmetz(2001)]{eke2001} Eke, V. R., Navarro, J. F., \& Steinmetz, M., 2001, ApJ, 554, 114
\bibitem[Evrard \& Henry(1991)]{evrard1991} Evrard, A. E. \& Henry, J. P., 1991, ApJ, 383, 95
\bibitem[Fixsen et al.(1996)]{fixsen1996} Fixsen, D. J. et al. 1996, ApJ, 473, 576
\bibitem[Gladders et al.(2006)]{gladders2006} Gladders, M. D., Yee, H. K. C., Majumdar, S., Barrientos, L. F., Hoekstra, H., Hall, P. B., \& Infante, L., 2006, astro-ph/0603588
\bibitem[Haiman et al.(2005)]{haiman2005} Haiman, Z., et al., 2005, astro-ph/0507013
\bibitem[Haiman, Mohr, \& Holder(2001)]{haiman2001} Haiman, Z., Mohr, J. J., \& Holder, G. P., 2001, ApJ, 553, 545
\bibitem[Helsdon \& Ponman(2000)]{helsdon2000} Helsdon, S. F.,\& Ponman, T. J., 2000, MNRAS, 315, 356
\bibitem[Henry \& Arnaud(1991)]{henry1991} Henry, J. P. \& Arnaud, K. A., 1991, ApJ, 372, 410
\bibitem[Henry(2004)]{henry2004} Henry, J. P., 2004, ApJ, 609, 603
\bibitem[Kaiser(1986)]{kaiser1986} Kaiser, N., 1986, MNRAS, 222, 323
\bibitem[Kaiser(1991)]{kaiser1991} Kaiser, N., 1991, ApJ, 383, 104
\bibitem[Kay et al.(2004)]{kay2004} Kay, S. T., Thomas, P. A., Jenkins, A. J., \& Pearce, F. R., 2004, MNRAS, 355, 1091
\bibitem[Lloyd-Davies, Ponman, \& Cannon(2000)]{lloyddavies2000} Lloyd-Davies, E. J., Ponman, T. J., \& Cannon, D. B., 2000, MNRAS, 315, 689
\bibitem[Makino, Sasaki, \& Suto(1998)]{makino1998} Makino, N., Sasaki, S., \& Suto, Y., 1998, ApJ, 497, 555
\bibitem[Majumdar \& Mohr(2004)]{majumdar2004} Majumdar, S. \& Mohr, J. J. 2004, ApJ, 613, 41
\bibitem[Markevich(1998)]{markevich1998} Markevich, M. 1998, ApJ, 504, 27
\bibitem[McCarthy et al.(2003a)]{mccarthy2003a} McCarthy, I. G., Babul, A., Holder, G. P., \& Balogh, M. L. 2003a, 591, 515
\bibitem[McCarthy et al.(2003b)]{mccarthy2003b} McCarthy, I. G., Babul, A., Holder, G. P., \& Balogh, M. L. 2003b, 591, 526
\bibitem[McCarthy et al.(2004)]{mccarthy2004} McCarthy, I. G., Balogh, M. J., Babul, A., Poole, G. B., \& Horner, D. J., 2004, ApJ, 613, 811
\bibitem[Motl et al.(2005)]{motl2005} Motl, P. M., Hallman, E. J., Burns, J. O., \& Norman, M. L., 2005, ApJ, 623, L63
\bibitem[Muanwong, Kay, \& Thomas(2006)]{muangwong2005} Muangwong, O., Kay, S. T., \& Thomas, P. A., 2006, ApJ, 649, 640
\bibitem[Nagai(2006)]{nagai2006} Nagai, D., 2006, ApJ, 650, 538 
\bibitem[Navarro, Frenk \& White(1995)]{nfw95} Navarro, J. F., Frenk, C. S., \& White, S. D. M., 1995, MNRAS, 275, 720
\bibitem[Navarro, Frenk \& White(1997)]{nfw97} Navarro, J. F., Frenk, C. S. \& White, S. D. M., 1997, ApJ, 490, 493 
\bibitem[Norman \& Bryan(1999)]{norman1999} Norman M. L. \& Bryan G. L., 1999, in Miyama S. M., Tomisaka K., Hanawa K., eds, Astrophys. Sci. Sci. Library Vol. 240, Numerical Astrophysics, Kluwer, Boston, p. 19
\bibitem[O'Shea et al.(2004)]{oshea2004} O'Shea, B. W., Bryan, G. L., Bordner, J., Norman, M. L., Abel, T., Harkness, R., \& Kritsuk, A., 2004, astro-ph/0403044
\bibitem[O'Shea et al.(2005)]{oshea2005} O'Shea, B. W., Nagamine, K., Springel, V., Hernquist, L., \& Norman, M. L., 2005, ApJS, 160, 1
\bibitem[Ostriker et al.(2005)]{ostriker2005} Ostriker, J.~P., Bode, P., \& Babul, A.\ 2005, \apj, 634, 964 
\bibitem[Pen(1999)] {pen1999} Pen, U., 1999, ApJ, 510, L1
\bibitem[Pierpaoli et al.(2003)]{pierpaoli2003} Pierpaoli, E., Borgani, S., Scott, D., \& White, M., 2003, MNRAS, 342, 163
\bibitem[Ponman et al.(1999)]{ponman1999} Ponman, T. J., Cannon, D. B., \& Navarro, J. F. 1999, Nature, 397, 135
\bibitem[Ponman, Sanderson, \& Finoguenov(2003)]{ponman2003} Ponman, T. J., Sanderson, A. J. R., \& Finoguenov, A., 2003, MNRAS, 343, 331
\bibitem[Pratt \& Arnaud(2003)]{pratt2003} Pratt, G. W. \& Arnaud, M., 2003, A\&A, 408, 1
\bibitem[Pratt \& Arnaud(2005)]{pratt2005} Pratt, G. W. \& Arnaud, M., 2005, A\&A, 429, 791
\bibitem[Pratt, Arnaud, \& Pointecouteau(2006)]{pratt2006} Pratt, G. W., Arnaud, M., \& Pointecouteau, E., 2006, A\&A, 446, 429
\bibitem[Raymond \& Smith(1977)]{raymondsmith1977} Raymond, J. C. \& Smith, B. W. 1977, ApJS, 35, 419
\bibitem[Reiprich \& B\"{o}hringer(2002)] {reiprich2002} Reiprich, T. \& B\"{o}hringer, H, 2002, ApJ, 567, 716
\bibitem[Rosati, Borgani, \& Norman(2002)]{rosati2002} Rosati, P., Borgani, S., \& Norman, C., 2002, AAR\&A, 40, 539
\bibitem[Shuecker et al.(2003)]{shuecker2003} Shuecker, P., B\"{o}hringer, H., Collins, C. A., \& Guzzo, L., 2003, A\&A, 298, 867
\bibitem[Spergel et al.(2006)]{spergel2006} Spergel, D. N., Bean, R., Dore', O., Nolta, M. R., Bennett, C. L., Hinshaw, G., Jarosik, N., Komatsu, E., Page, L., Peiris, H. V., Verde, L., Barnes, C., Halpern, M., Hill, R. S., Kogut, A., Limon, M., Meyer, S. S., Odegard, N., Tucker, G. S., Weiland, J. L., Wollack, E., \& Wright, E. L., ApJ, submitted, astro-ph/0603449 
\bibitem[Springel(2005)]{gadget2}Springel, V., 2005, MNRAS, 364, 1105
\bibitem[Sunyaev \& Zeldovich(1972)]{sunyaev1972} Sunyaev, R. \& Zeldovich, Y., 1972, Comments Aprophys. Space Phys., 2, 66
\bibitem[Sunyaev \& Zeldovich(1980)]{sunyaev1980} Sunyaev, R. \& Zeldovich, Y., 1980, MNRAS, 190, 413
\bibitem[Tozzi \& Norman(2001)]{tozzi2001} Tozzi, P. \& Norman, C., 2001, ApJ, 546, 63
\bibitem[Viana \& Liddle(1996)]{viana1996} Viana, P. P. \& Liddle, A. R., 1996, MNRAS, 262, 1023
\bibitem[Voit \& Bryan(2001)]{voit2001} Voit, M. G. \& Bryan, G. L., 2001, ApJ, 414, 425
\bibitem[Voit et al.(2002)]{voit2002} Voit, M. G, Bryan, G. L., Balogh, M.  L., \& Bower, R. G., 2002, ApJ, 576, 601
\bibitem[Voit et al.(2003)]{voit2003} Voit, M. G., Balogh, M. L., Bower, R. G., Lacey, C. G., \& Bryan, G. L., 2003, ApJ, 593, 272
\bibitem[Voit \& Ponman(2003)]{voitponman2003} Voit, M. G. \& Ponman, T. J., 2003, ApJL, 594, 75
\bibitem[Voit(2005)]{voit2005} Voit, G. M., 2005, Rev. Mod. Phys., 77, 207
\bibitem[Voit, Kay, \& Bryan(2005)]{voitkaybryan2005} Voit, M. G., Kay, S. T., \& Bryan, G. L., 2005, MNRAS, 364, 909
\bibitem[Wang et al.(2004)]{wang2004} Wang, S., Khoury, J., Haiman, Z., \& May, M. 2004, Phys. Rev. D., 70, 7013008 
\bibitem[Wu, Fabian, \& Nulsen(2001)]{wu2001} Wu, K. K. S., Fabian, A. C., \& Nulsen, P. E. J., 2001, MNRAS, 324, 95
\bibitem[Younger, Bahcall, \& Bode(2005)]{younger2005} Younger, J. D., Bahcall, N. A., \& Bode, P., 2005, ApJ, 622, 1 
\bibitem[Younger et al.(2006)]{younger2006} Younger, J. D., Haiman, Z., Bryan, G. L., \& Wang, S., 2006, ApJ, 653, 27

\end{thebibliography}
\end{document}